\documentstyle[11pt,newpasp,twoside,epsf]{article}
\def \lumstar{$L_B=10^{10}h^{-2} {\rm L}_{\rm {B}\odot}~$}

\def\edcomment#1{\iffalse\marginpar{\raggedright\sl#1\/}\else\relax\fi}
\reversemarginpar

\begin{document}
\title{Properties of galaxy dark matter halos from weak lensing}
 \author{Henk Hoekstra}
\affil{Canadian Institute for Theoretical Astrophysics, University of
Toronto, 60 St. George Street, M5S 3H8, Toronto, Canada}
\author{Howard K.C. Yee}
\affil{Department of Astronomy and Astrophysics, University of
Toronto, 60 St. George Street, M5S 3H8, Toronto, Canada}
\author{Michael D. Gladders}
\affil{Observatories of the Carnegie Institution of Washington, 813 Santa
Barbara Street, Pasadena, CA 91101}

\begin{abstract}

We present the results of a study of the average mass profile around
galaxies using weak gravitational lensing. We use 45.5 deg$^2$ of
$R_C$ band imaging data from the Red-Sequence Cluster Survey (RCS) and
define a sample of $\sim 1.2\times 10^5$ lenses with $19.5<R_C<21$,
and a sample of $\sim 1.5\times 10^6$ background galaxies with
$21.5<R<24$. 

 We constrain the power law scaling relations between the $B$-band
luminosity and the mass and size of the halo, and find that the
results are in excellent agreement with observed luminosity-line-width
relations. Under the assumption that the luminosity does not evolve
with redshift, the best fit NFW model yields a mass
$M_{200}=(8.8\pm0.7)\times 10^{11} h^{-1} M_\odot$ and a scale radius
$r_s=16.7^{+3.7}_{-3.0} h^{-1}$ kpc for a galaxy with a fiducial
luminosity of \lumstar. The latter result is in excellent agreement
with predictions from numerical simulations for a halo of this
mass. We also observe a signficant anisotropy of the lensing signal
around the lenses, implying that the halos are flattened and aligned
with the light distribution. We find an average (projected) halo
ellipticity of $\langle e_{\rm halo} \rangle=0.20^{+0.04}_{-0.05}$, in
fair agreement with results from numerical simulations of
CDM. Alternative theories of gravity (without dark matter) predict an
isotropic lensing signal, which is excluded with 99.5\%
confidence. Hence, our results provide strong support for the
existence of dark matter.

\end{abstract}

\section{Introduction}

The existence of massive dark matter halos around galaxies is widely
accepted, based on different lines of evidence, such as flat rotation
curves of spiral galaxies and strong lensing systems. However,
relatively little is known about the properties of dark matter halos,
such as their extent and shapes.

A promising approach to study the galaxy dark matter halos is weak
gravitational lensing. The tidal gravitational field of the dark
matter halo introduces small coherent distortions in the images of
distant background galaxies. The weak lensing signal can be measured
out to large projected distances from the lens, and hence provides a
unique probe of the gravitational potential on large scales.  The
applications of this approach are numerous: one can infer masses of
galaxies and compare the results to their luminosities (e.g., McKay et
al. 2001), or one can attempt to constrain the halo mass profile
(e.g., Fischer et al. 2000; McKay et al. 2001). Also, weak lensing can
be used to constrain the shapes of halos by measuring the azimuthal
variation of the lensing signal. 

Unfortunately, one can only study ensemble averaged properties,
because the weak lensing signal induced by an individual galaxy is too
low to be detected. Nevertheless, as we will show here, weak lensing
is a useful probe of the matter distribution in galaxies, and we
expect it to make significant contributions to our understanding of
galaxy formation in the coming years.

In these proceedings we highlight the main results from our weak
lensing analysis. A detailed description of our results can be found
in Hoekstra, Yee, \& Gladders (2003).

\section{Data \& Analysis}

We use 45.5 deg$^2$ of $R_C$-band imaging data from the Red-Sequence
Cluster Survey (e.g., Yee \& Gladders 2001), which were taken with the
CFH12k camera on the CFHT. A detailed description of the data
reduction and object analysis can be found in Hoekstra et al. (2002),
to which we refer for technical details.

For the analysis presented here, we select a sample of ``lenses'' and
``sources'' on the basis of their apparent $R_C$ magnitude. We define
galaxies with $19.5<R_C<21$ as lenses, and galaxies with $21.5<R_C<24$
as sources which are used to measure the lensing signal. This
selection yields a sample of $\sim 1.2\times 10^5$ lenses and $\sim
1.5\times 10^6$ sources.

The interpretation of the weak lensing signal (e.g., inferring sizes
and masses for the galaxy halos), requires knowledge of the redshift
distributions of both lenses and sources. The redshift distribution of
the sample of lenses has been determined spectroscopically by the
CNOC2 Field Galaxy Redshift Survey (e.g., Yee et al. 2000). The
derived redshift distribution gives a median redshift $z=0.35$ for the
lens sample. We use the redshifts and the colors of the galaxies
observed in the CNOC2 survey to compute their rest-frame $B$
luminosity.  For the source galaxies the situation is more
complicated.  These galaxies are generally too faint for spectroscopic
surveys. Instead we use the photometric redshift distributions derived
from both Hubble Deep Fields, which yields a median redshift of
$z=0.53$ for the source galaxies.

Weak lensing measures the convolution of the galaxy distribution and
the galaxy dark matter profiles. To examine the ensemble average
properties of the dark matter halos properly, we need to account for
the clustering of the lenses.  This is done naturally in a maximum
likelihood analysis, where a model for the mass distribution of
individual galaxies is compared to the observations. We use the
profile predicted by cold dark matter simulations (e.g., Navarro,
Frenk, \& White 1995, NFW hereafter).  The NFW density profile is
characterized by two parameters, a density contrast $\delta_c$ and a
scale $r_s$

\begin{equation}
\rho(r)=\frac{\delta_c\rho_c}{(r/r_s)(1+r/r_s)^2},
\end{equation}

\noindent where $\rho_c$ is the critical surface density at the
redshift of the halo. The ``virial'' radius $r_{200}$ is defined as
the radius where the mass density of the halo is equal to
$200\rho_c$. The virial mass $M_{200}$ is defined as the mass enclosed
within $r_{200}$, with a corresponding rotation velocity
$V_{200}=V_c(r_{200})$.

In these proceedings we compare the NFW profile to the observations,
with $V_{200}$ (or equivalently $M_{200}$) and $r_s$ as free
parameters.  Hoekstra et al. (2003) also consider a truncated
isothermal sphere (TIS) model.

To infer the best estimates for the model parameters, one formally has
to perform a maximum likelihood analysis in which the redshift of each
individual galaxy is a free parameter, which has to be chosen such
that it maximizes the likelihood. This approach is computationally not
feasible, and instead we create mock redshift catalogs, using the
observed redshift distributions from the CNOC2 survey (see Hoekstra et
al.  2003 for details), which allows us to find estimates for the
model parameters, close to the true maximum likelihood values.

\begin{figure}[!h]
\begin{center}
\epsfxsize=0.7\hsize
\epsfbox{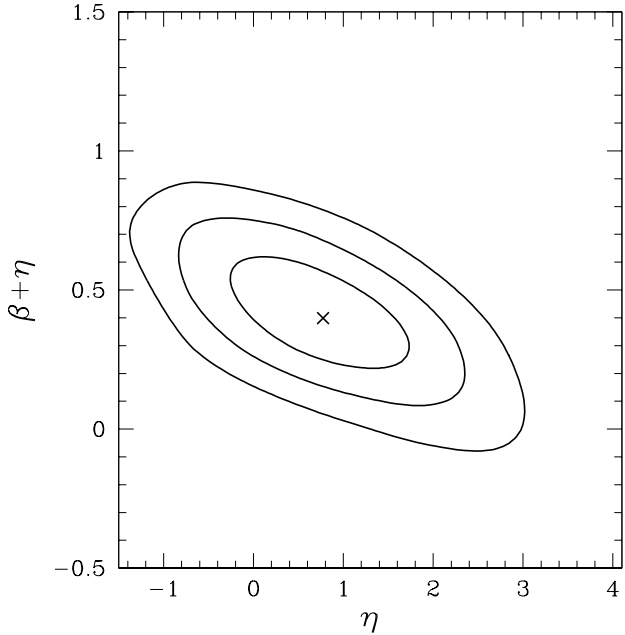}
\end{center}
\vspace{-0.5cm}
\caption{\small Joint constraints on $\beta+\eta$ and $\eta$
from the maximum likelihood analysis using an NFW model.  The cross
indicates the best estimates for $\beta+\eta$ and $\eta$.  The
contours indicate the 68.3\%, 95.4\%, and the 99.7\% confidence on two
parameters jointly. The mass $M_{200}$ scales approximately with
luminosity $\propto L_{\rm B}^{1.5(\beta+\eta)}$.
\label{scaling_nfw}}
\end{figure}

\section{Scaling relations}

We first examine how the model parameters $\delta_c$ and $r_s$ scale
with luminosity. We assume that the scaling relations have a
power law dependence on the restframe $B$-band luminosity:

\begin{equation}
\delta_c\propto L_{\rm B}^{\beta},{\rm~and~}
r_s\propto L_{\rm B}^{\eta/2},
\end{equation}

Currently we use the apparent $R_C$-band magnitude as a crude measure
of the luminosity. With upcoming multi-color data, we expect to
significantly improve on our measurements. To derive joint constraints
on $\beta$ and $\eta$, we marginalise over the values for $\delta_c$
and $r_s$. It is more convenient to constrain $\beta+\eta$ and $\eta$,
and the results are presented in Figure~\ref{scaling_nfw}.

Figure~\ref{scaling_nfw} shows that the sum $\beta+\eta$ is fairly
well constrained. We obtain a best fit value of
$\beta+\eta=0.40^{+0.14}_{-0.13}$ (68\% confidence). For $\eta$ we
find $\eta=0.77^{+0.63}_{-0.67}$.  Also the relation between the
maximum rotation velocity $V_{\rm max}$ and the luminosity is well
described by a power law, with an exponent $0.20^{+0.13}_{-0.14}$
(68\% confidence) in good agreement with the observed slope of the
Tully-Fisher relation (e.g., Verheijen 2001).

\section{Mass and extent of halos}

Many different lines of evidence suggest that galaxies are surrounded
by massive dark matter halos, but observationally it is difficult to
place constraints on the mass and extent of these halos, because of
the lack of visible tracers that can be used to infer the
gravitational potential. Weak gravitational lensing does not suffer
from the latter requirement and consequently provides one of the most
powerful measures of the mass distribution at large radii. With the
current data it is not possible to distinguish between different mass
profiles (as mentioned above we use the NFW profile), but for a given
model, we can examine the masses and sizes of the halos.

\begin{figure}[!t]
\epsfxsize=\hsize
\epsfbox{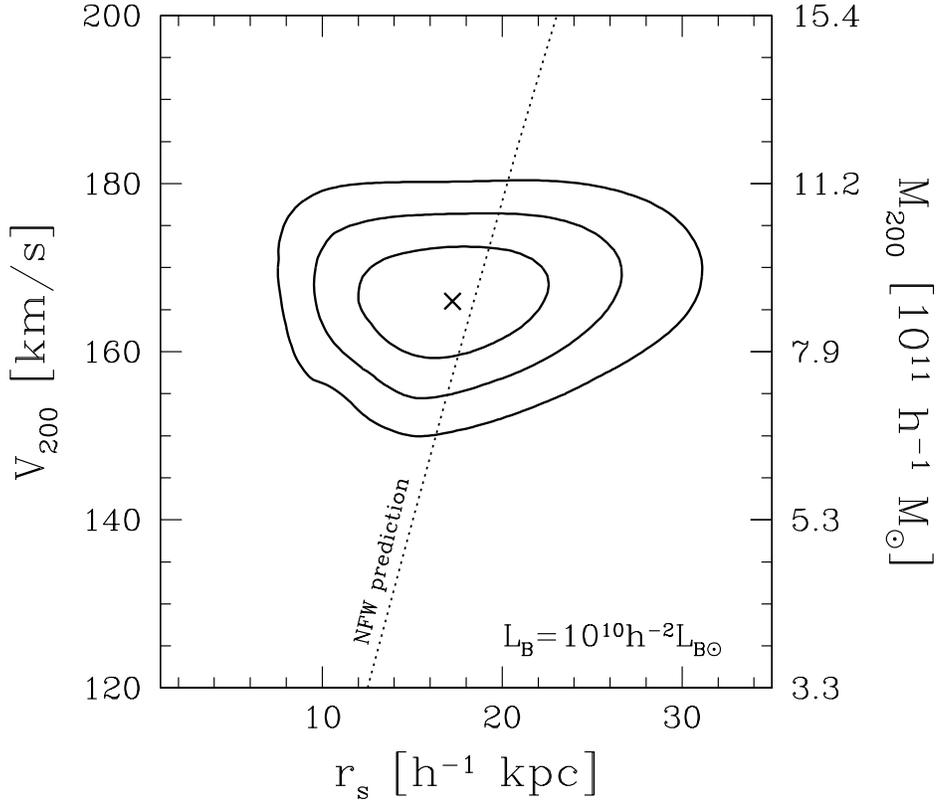}
\caption{\small Joint constraints on $V_{200}$ and scale radius
$r_s$ for a fiducial galaxy with $L_{\rm B}=10^{10}h^{-2}L_{{\rm
B}\odot}$, with an NFW profile. The corresponding values for $M_{200}$
are indicated on the right axis. The contours indicate the 68.3\%,
95.4\%, and the 99.7\% confidence on two parameters jointly. The cross
indicates the best fit value. The dotted line indicates the
predictions from the numerical simulations, which are in excellent
agreement with our results.
\label{size_nfw}}
\end{figure}

Figure~\ref{size_nfw} shows the joint constraints on $V_{200}$ (the
corresponding values for $M_{200}$ are indicated on the right) and
$r_s$ for a galaxy with a luminosity of \lumstar (under the assumption
that the luminosity does not evolve with redshift). The mass is
well constrained and we find a best fit value of $V_{200}=166\pm5$ km/s
or $M_{200}=(8.8\pm0.7)\times 10^{11} h^{-1} M_\odot$ (68\% confidence).
For the scale $r_s$ we find $r_s=16.7^{+3.7}_{-3.0} h^{-1}$ kpc (68\%
confidence).

In our maximum likelihood analysis we considered $r_s$ and $V_{200}$
free parameters. Numerical simulations, however, show that the
parameters in the NFW model are correlated, albeit with some
scatter. As a result, the NFW model can be considered as a
one-parameter model. The dotted line in Figure~\ref{size_nfw}
indicates this prediction. If the simulations provide a good
description of dark matter halos, the dotted line should intersect our
confidence region, which it does.

This result provides important support for the CDM paradigm, as the
latter predicts a ``size'' of dark matter halos which is in excellent
agreement with our observations. It is important to note that this
analysis is a direct test of CDM (albeit not conclusive), because the
weak lensing results are inferred from the gravitational potential at
large distances from the galaxy center, where dark matter dominates.
Most other attempts to test CDM are confined to the inner regions,
where baryons are, or might be, important.

\section{Shapes of halos}

The average shape of dark matter halos can provide important
information about the nature of dark matter. Numerical simulations of
cold dark matter yield triaxial halos, with a typical ellipticity of
$\sim 0.3$ (e.g., Dubinski \& Carlberg 1991).  Hence, in the context
of collisionless cold dark matter, the theoretical evidence for
flattened halos is quite strong. If the dark matter is interacting, it
tends to produce halos that are more spherical (compared to cold dark
matter). This difference is more pronounced in the central parts of
the halo, where the density is high.  On the large scales probed by
weak lensing, the different types of dark matter (for reasonable
interaction cross-sections) produce halos with similar shapes.

Weak gravitational lensing is potentially the most powerful way to
derive constraints on the shapes of dark matter halos.  The amount of
data required for such a measurement, however, is large: the
galaxy-galaxy lensing signal is small, and now one needs to measure an
even smaller azimuthal variation. We also have to assume that halo
is aligned with the galaxy. An imperfect alignment between light
and halo will reduce the amplitude of the azimuthal variation
detectable in the weak lensing analysis.  Hence, weak lensing formally 
provides a lower limit to the average halo ellipticity.

\begin{figure}
\begin{center}
\epsfxsize=0.7\hsize
\epsfbox{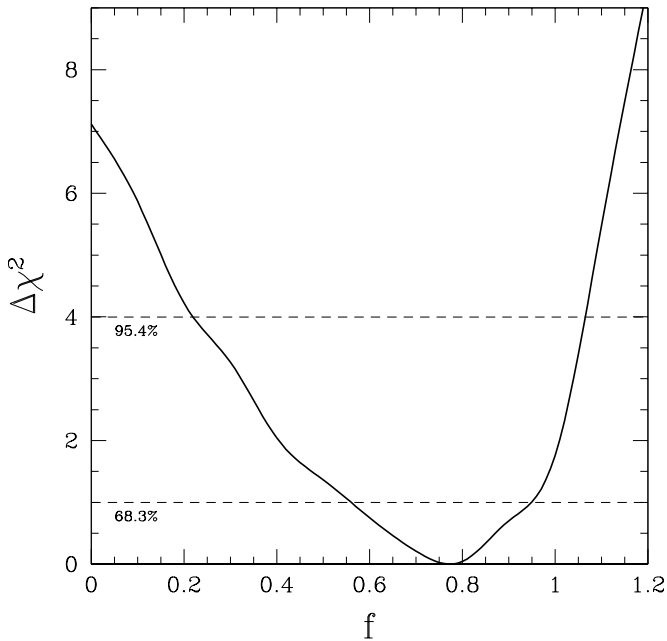}
\end{center}
\caption{\small $\Delta \chi^2$ as a function of $f$. We have
assumed that the ellipticity of the halos is related to the observed
ellipticity of the lens as $e_{\rm halo}=f e_{\rm lens}$.  We have
indicated the 68.3\% and 95.4\% confidence intervals. We find a best
fit value of $f=0.77^{+0.18}_{-0.21}$ (68\% confidence). Round halos
$(f=0)$ are excluded with 99.5\% confidence. 
\label{flat_halo}}
\end{figure}

To maximize the signal-to-noise ratio of the measurement one has to
assign proper weights to the lenses: edge-on galaxies have maximal
weight, whereas the lensing signal around face-on galaxies contains no
information about the shape of the halo. We adopt a simple approach,
and assume that the (projected) ellipticity of the dark matter halo is
proportional to the shape of the galaxy: $e_{\rm halo}=f e_{\rm
lens}$.

As before, we compute the model shear field, and compare this to the
data. Figure~\ref{flat_halo} shows the resulting $\Delta\chi^2$ as a
function of $f$.  We find a best fit value of $f=0.77^{+0.18}_{-0.21}$
(68\% confidence). This suggests that, on average, the dark matter
distribution is rounder than the light distribution. As discussed
above, our analysis formally provides only a lower limit on the halo
ellipticity, and the true ellipticity might be higher if some of the
halos are misaligned with the light.  Nevertheless, the fact that we
detect a significant flattening implies that the halos are well
aligned with the light distribution. Also note that the lensing signal
is caused by a range of different galaxy types, for which our simple
relation between the halo ellipticity and light distribution might not
be valid.

A simple interpretation of the results is difficult, but a simple
approach actually yields sensible results. For instance, the average
ellipticity of the lens galaxies is $\langle e_{\rm
lens}\rangle=0.261$. Hence, the measured value of $f$ implies an
average projected halo ellipticity of $\langle e_{\rm
halo}\rangle=0.20^{+0.04}_{-0.05}$ (68\% confidence), which
corresponds to an projected axis ratio of $c/a=0.66^{+0.07}_{-0.06}$
(68\% confidence). Although the weak lensing yields a projected axis
ratio, the result is in fair agreement with the results from numerical
simulations.

A robust outcome of our analysis is that spherical halos ($f=0$) are
excluded with 99.5\% confidence. This result poses a serious problem
for alternative theories of gravity, which attempt to explain the
observations without the need of dark matter. In such theories the
lensing signal we measure is effectively caused by point masses (the
visible matter is confined to much smaller scales). As a result, in
such theories one would expect an almost isotropic lensing signal
around the galaxies (even if the quadrupole decreases as slow as
$r^{-\sqrt{3}}$, e.g., Binney this proceedings), which is not
observed. Note that this test does not require knowledge of the actual
deflection angles, which typically are unknown for these
theories. Hence, our results provide strong support for the existence
of dark matter.

\end{document}